\documentclass{article}%
\usepackage{amsmath}
\usepackage{amsfonts}
\usepackage{amssymb}
\usepackage{graphicx}%
\newcommand{\bpartial}{\mathop{\partial\kern -4pt\raisebox{.8pt}{$|$}}}
\newcommand{\bra}{\mathopen{[\kern-1.6pt[}}
\newcommand{\ket}{\mathclose{]\kern-1.5pt]}}
\newcommand{\bbra}{\mathopen{[\kern-2.2pt[\kern-2.3pt[}}
\newcommand{\bket}{\mathclose{]\kern-2.1pt]\kern-2.3pt]}}

\makeindex
\begin{document}

\title { Classification of two and three dimensional Lie super-bialgebras}
\vspace{3mm}

\author {  A. Eghbali \hspace{-2mm}{ \footnote{ e-mail: a.eghbali@azaruniv.edu}}\hspace{1mm},\hspace{1mm} A. Rezaei-Aghdam
\hspace{-2mm}{ \footnote{Corresponding author. e-mail:
rezaei-a@azaruniv.edu}}\hspace{2mm}{\small and}
F. Heidarpour \\
{\small{\em Department of Physics, Faculty of science, Azarbaijan University }}\\
{\small{\em of Tarbiat Moallem , 53714-161, Tabriz, Iran  }}}

\maketitle

\vspace{3cm}

\begin{abstract}
Using adjoint representation of Lie superalgebras, we obtain the
matrix form of super Jacobi and mixed super Jacobi identities of
Lie super-bialgebras. By direct calculations of these identities,
and use of automorphism supergroups of two and three dimensional
Lie superalgebras, we obtain and classify all two and three
dimensional Lie super-bialgebras.
\end{abstract}

\newpage
\section{\bf Introduction}
Lie bialgebras were first introduced, from the mathematical point
of view, by Drinfel'd as algebraic structures and classical limit
of underlying quantized enveloping algebras ({\it quantum
groups}) \cite{Drin}. In particular, every deformation of a
universal enveloping algebra induces a Lie bialgebra structure on
the underlying Lie algebra. Conversely, it has been shown in Ref.
\cite{Etin} that each Lie bialgebra admits quantization. So the
classification of Lie bialgebras can be seen as the first step in
the classification of quantum groups. Many interesting examples of
Lie bialgebras based on complex semisimple Lie algebras have been
given in Drinfel'd \cite{Drin}. A complete classification of Lie
bialgebras with reduction was given in Ref. \cite{Del}. However, a
classification of Lie bialgebras is out of reach, with similar
reasons as for Lie algebra classification. In the non-semisimple
case, only a bunch of low dimensional examples have been
thoroughly studied \cite{{JR},{Gomez},{Snobl},{RHR}}. On the
other hand, from the physical point of view, the theory of
classical integrable systems naturally relates to the geometry and
representation theory of Poisson-Lie groups and the corresponding
Lie bialgebras and their classical r-matrices (see, for example,
Ref. \cite{Kosmann}). In the same way, Lie super-bialgebras
\cite{N.A}, as the underlying symmetry algebras, play an
important role in the integrable structure of $AdS/CFT$
correspondence \cite{Bs}. In this way, and by considering that
there is a universal quantization for Lie super-bialgebras
\cite{Geer}, one can assign an important role to the
classification of Lie super-bialgebras (especially low
dimensional Lie super-bialgebras) from both  physical and
mathematical point of view. There are distinguished and
nonsystematic ways for obtaining low dimensional Lie
super-bialgebras (see, for example, Refs.
\cite{{J.z},{J},{KARAALI}}). In this paper, using the adjoint
representation of Lie superalgebras, we present a systematic way
for obtaining and classifying  low dimensional Lie
super-bialgebras. We apply this method to the classification of
two and three dimensional Lie super-bialgebras.

\smallskip
\goodbreak

The paper is organized as follows. In section two, we give basic
definitions and notations that are used throughout the paper. The
systematic way for classification of Lie super-bialgebras by using
matrix form of super Jacobi $(sJ)$ and mixed super Jacobi $(msJ)$
identities of Lie super-bialgebras is described in section three.
A list of two and three dimensional Lie superalgebras of Ref.
\cite{B} is offered in section 4. The automorphism Lie
supergroups of these Lie superalgebras are also presented in
section 4 . Then, using the method mentioned in section two, we
classify all $48$ two and three dimensional Lie super-bialgebras
in section five. The details of calculations are explained, using
an example. The concluding section discusses some remarks.
Certain properties of tensors and supermatrices are given in
appendix A. Solutions of $(sJ-msJ)$ identities and their
isomorphism matrices are give in appendix B.

\vspace{7mm}

\section{\bf Basic definitions and notations }

$\; \; \; \;$In this paper we use DeWitt notation for supervector
spaces, supermatrices, etc \cite{D}. Some of definitions and
related notations are given in appendix A.

\bigskip

{\it Definition}: A {\em Lie superalgebra} ${\bf g}$ is a graded
vector space ${\bf g}={\bf g}_B \oplus {\bf g}_F$ with gradings;
$grade({\bf g}_B)=0,\; grade({\bf g}_F)=1$; such that Lie bracket
satisfies the super antisymmetric and super Jacobi identities,
i.e., in a graded basis $\{X_i\}$ of ${\bf g}$ if we put \cite{F1}

\begin{equation}
[X_i , X_j] = {f^k}_{ij} X_k,
\end{equation}
then
\begin{equation}
(-1)^{i(j+k)}{f^m}_{jl}{f^l}_{ki} + {f^m}_{il}{f^l}_{jk} +
(-1)^{k(i+j)}{f^m}_{kl}{f^l}_{ij}=0,
\end{equation}
so that
\begin{equation}
{f^k}_{ij}=-(-1)^{ij}{f^k}_{ji}.
\end{equation}
Note that, in the standard basis, ${f^B}_{BB}$ and ${f^F}_{BF}$
are real c-numbers and ${f^B}_{FF}$ are pure imaginary c-numbers
and other components of structure constants ${f^i}_{jk}$ are zero
\cite{D}, i.e. we have
\begin{equation}
{f^k}_{ij}=0, \hspace{10mm} if \hspace{5mm} grade(i) +
grade(j)\neq grade(k)\hspace{3mm} (mod 2).
\end{equation}
Let ${\bf g}$ be a finite-dimensional Lie superalgebra and ${\bf
g}^\ast$ be its dual vector space and let $( .~, ~. )$ be the
canonical pairing on ${\bf g}^\ast \oplus {\bf g}$.

\bigskip

{\it Definition}: A {\em Lie super-bialgebra} structure on a Lie
superalgebra ${\bf g}$ is a super skew-symmetric linear map
$\delta : {\bf g }\longrightarrow {\bf g}\otimes{\bf g}$ ({\em the
super cocommutator}) so that \cite{N.A}

1)~$\delta$ is a super one-cocycle, i.e.
\begin{equation}
\delta([X,Y])=(ad_X\otimes I+I\otimes
ad_X)\delta(Y)-(-1)^{|X||Y|}(ad_Y\otimes I+I\otimes
ad_Y)\delta(X) \qquad \forall X,Y\in {\bf g},
\end{equation}
~~~~~~~~where $|X|(|Y|)$ indicates the grading of $X(Y)$;

2) the dual map ${^t}{\delta}:{\bf g}^\ast\otimes {\bf g}^\ast \to
{\bf g}^\ast$ is a Lie superbracket on ${\bf g}^\ast$, i.e.,
\begin{equation}
(\xi\otimes\eta , \delta(X)) = ({^t}{\delta}(\xi\otimes\eta) , X)
= ([\xi,\eta]_\ast , X)    \qquad \forall X\in  {\bf g} ;\;\,
\xi,\eta\in{\bf g}^\ast.
\end{equation}
The Lie super-bialgebra defined in this way will be indicated by
$({\bf g},{\bf g}^\ast)$ or $({\bf g},\delta)$.

\bigskip

 {\it Proposition}: If there exists an automorphism $A$ of
${\bf g}$ such that
\begin{equation}
\delta^\prime = (A\otimes A)\circ\delta\circ A^{-1},
\end{equation}
then the super one-cocycles $\delta$ and $\delta^\prime$ of the
Lie superalgebra $\bf g$ are {\em equivalent} \cite{F2}. In this
case the two Lie super-bialgebras $({\bf g},\delta)$ and $({\bf
g},\delta^\prime)$ are equivalent (as in the bosonic case
\cite{RHR}).

\bigskip

{\it Definition}: A {\em Manin} super triple \cite{N.A} is a
triple of Lie superalgebras $(\cal{D} , {\bf g} , {\bf
\tilde{g}})$ together with a non-degenerate ad-invariant super
symmetric bilinear form $<.~, ~. >$ on $\cal{D}$ such
that\hspace{2mm}

1)$\;$${\bf g}$ and ${\bf \tilde{g}}$ are Lie sub-superalgebras of
$\cal{D}$,\hspace{2mm}

2)$\;$ $\cal{D} = {\bf g}\oplus{\bf \tilde{g}}$ as a supervector
space,\hspace{2mm}

3)$\;$ ${\bf g}$ and ${\bf \tilde{g}}$ are isotropic with respect
to $<  .~, ~. >$, i.e.
\begin{equation}
<X_i , X_j> = <\tilde{X}^i , \tilde{X}^j> = 0, \hspace{10mm}
{\delta_i}^j=<X_i , \tilde{X}^j> = (-1)^{ij}<\tilde{X}^j,
X_i>=(-1)^{ij}{\delta^j}_i,
\end{equation}
where $\{X_i\}$ and $\{\tilde{X}^i\}$ are  basis of Lie
superalgebras ${\bf g}$ and ${\bf \tilde{g}}$, respectively. Note
that in the above relation ${\delta^j}_i$ is the ordinary delta
function. There is a one-to-one correspondence between Lie
super-bialgebra $({\bf g},{\bf g}^\ast)$ and Manin super triple
$(\cal{D} , {\bf g} , {\bf \tilde{g}})$ with $ {\bf g}^\ast={\bf
\tilde{g}}$ \cite{N.A}. If we choose the structure constants of
Lie superalgebras ${\bf g}$ and ${\bf \tilde{g}}$ as
\begin{equation}
[X_i , X_j] = {f^k}_{ij} X_k,\hspace{20mm} [\tilde{X}^i ,\tilde{ X}^j] ={{\tilde{f}}^{ij}}_{\; \; \: k} {\tilde{X}^k}, \\
\end{equation}
then ad-invariance of the bilinear form $<.~ ,~. >$ on $\cal{D} =
{\bf g}\oplus{\bf \tilde{g}}$ implies that
\begin{equation}
[X_i , \tilde{X}^j] =(-1)^j{\tilde{f}^{jk}}_{\; \; \; \:i} X_k
+(-1)^i {f^j}_{ki} \tilde{X}^k.
\end{equation}
Clearly, using  Eqs. (6) and (9) we have
\begin{equation}
\delta(X_i) = (-1)^{jk}{\tilde{f}^{jk}}_{\; \; \; \:i} X_j \otimes
X_k,
\end{equation}
note that the appearance of $(-1)^{jk}$ in this relation is due to
the definition of natural inner product between ${\bf g}\otimes
{\bf g}$ and  ${\bf g}^\ast \otimes {\bf g}^\ast $ as $
(\tilde{X}^i \otimes \tilde{ X}^j ,X_k \otimes
X_l)=(-1)^{jk}{\delta^i}_k {\delta^j}_l$.

As a result, if we apply this relation in the super one-cocycle
condition (5), super Jacobi identities (2) for the  dual Lie
superalgabra and the following mixed super Jacobi identities are
obtained
\begin{equation}
{f^m}_{jk}{\tilde{f}^{il}}_{\; \; \; \; m}=
{f^i}_{mk}{\tilde{f}^{ml}}_{\; \; \; \; \; j} +
{f^l}_{jm}{\tilde{f}^{im}}_{\; \; \; \; \; k}+ (-1)^{jl}
{f^i}_{jm}{\tilde{f}^{ml}}_{\; \; \; \; \; k}+ (-1)^{ik}
{f^l}_{mk}{\tilde{f}^{im}}_{\; \; \; \; \; j}.
\end{equation}
This relation can also be obtained from super Jacobi identity of
$\cal{D}$.

\vspace{7mm}

\section{\bf Calculation of Lie super-bialgebras using adjoint representation }
$\; \; \; \;$As discussed in the introduction, by use of super
Jacobi and mixed super Jacobi identities we  classify two and
three dimensional Lie super-bialgebras. This method was first
used to obtain three dimensional Lie bialgebras in Ref. \cite{JR}.
Because of tensorial form of super Jacobi and  mixed super Jacobi
identities, working with them is not so easy and we suggest
writing these equations as matrix forms using the following
adjoint representations for Lie superalgebras ${\bf g}$ and ${\bf
\tilde{g}}$
\begin{equation}
({\tilde{\cal X}}^i)^j_{\; \;k} = -{\tilde{f}^{ij}}_{\; \; \; k},
\hspace{10mm} ({\cal Y}^i)_{ \;jk} = -{f^i}_{jk}.
\end{equation}
Then the matrix forms of super Jacobi identities (2) for dual Lie
superalgebra $\tilde {\bf g}$ and mixed super Jacobi identities
(12) become as follows, respectively:
\begin{equation}
({\tilde{\cal X}}^i)^j_{\; \;k}{\tilde{\cal X}}^k - {\tilde{\cal
X}}^j {\tilde{\cal X}}^i + (-1)^{ij}{\tilde{\cal X}}^i
{\tilde{\cal X}}^j = 0,
\end{equation}
\begin{equation}
{({\tilde{\cal X}}^i)}^j_{\; \;l}\;{\cal Y}^l =-(-1)^{k}
({\tilde{\cal X}}^{st})^{j}\; {\cal Y}^i + {\cal Y}^j{\tilde{\cal
X}}^i - (-1)^{ij}{\cal Y}^i {\tilde{\cal X}}^j + (-1)^{k+ij}
({\tilde{\cal X}}^{st})^{i} \;{\cal Y}^j.
\end{equation}
Note that in the above relations we use the right indices for the
matrix elements and index $k$ represents the column of matrix
${\tilde{\cal X}}^{st}$.

Having the structure constants of the Lie superalgebra ${\bf g}$,
we solve the matrix Eqs. (14) and (15) in order to obtain the
structure constants of the dual Lie superalgebras ${\bf {\tilde
g}}$ so that $({\bf g}, {\bf {\tilde g}})$ is Lie super-bialgebra.
We assume that, we have classification and tabulation of the Lie
superalgebras ${\bf g}$ (for example two and three dimensional Lie
superalgebras for this paper). Here we explain the method in
general and one can apply it for classification of other low
dimensional Lie super-bialgebras \cite{RE}. We fulfill this work
in the following three steps. Note that in solving Eqs. (14),
(15), (17) and (21) we use the  MAPLE program.

\bigskip

{\it  Step 1:}~~{\it Solutions of super Jacobi and mixed super
Jacobi identities and determination of Lie superalgebras $\bf g'$
which are isomorphic with dual solutions}

\smallskip

With the solution of matrix Eqs. (14) and (15) for obtaining
matrices ${\tilde{\cal X}}^i$, some structure constants of ${\bf
{\tilde g}}$ are obtained to be zero, some unknown and some
obtained in terms of each other. In order to know whether ${\bf
{\tilde g}}$ is one of the Lie superalgebras of table or
isomorphic to them, we must use the following isomorphic relation
between obtained Lie superalgebras ${\bf {\tilde g}}$ and one of
the Lie superalgebras of table, e.g., $\bf g'$. Applying the
transformation (37) for a change of  basis ${\bf {\tilde g}}$ we
have
\begin{equation}
\tilde{X}^{'\;i}=C^i_{\;\;j}\tilde{X}^j,\hspace{20mm}
[\tilde{X}^{'\;i} ,\tilde{X}^{'\;j}] ={\tilde{f}^{'\;ij}}_{\; \;
k} \tilde{X}^{'\;k},
\end{equation}
then  the following matrix equations for isomorphism
\begin{equation}
(-1)^{i(j+l)}\; C\;(C^i_{\;\;k}\;\tilde{\cal X}^k_{\tilde{(\bf
g)}})={\cal X}^{i}_{(\bf g')}\;C,
\end{equation}
are obtained, where the indices $j$ and $l$ correspond to the row
and column of  $C$ in the left hand side of (17). In the above
matrix equations ${\cal X}^{i}_{(\bf g')}$ are adjoint matrices
of known Lie superalgebra $\bf g'$ of the classification table.
Solving (17) with the condition $sdet C\neq 0$ we obtain some
extra conditions on ${{\tilde{f}^{kl}}_{(\bf {\tilde
g})}}\hspace{1mm}_m$'s that were obtained from (14) and (15).

\bigskip

{\it Step 2:}~~{\it Obtaining general form of the transformation
matrices $B:{\bf g}'\longrightarrow {\bf g}'.i$; such that $({\bf
g}, {\bf g}'.i)$ are Lie super-bialgebras}

\smallskip

As the second step we transform Lie super-bialgebra $({\bf g},
{\bf {\tilde g}})$ (where in the Lie superalgebra $\bf {\tilde
g}$ we impose extra conditions obtained in the step one) to Lie
super bialgebra  $({\bf g}, {\bf g}'.i)$ (where ${\bf g}'.i$ is
isomorphic as Lie superalgebra to ${\bf g}'$) with an
automorphism of the Lie superalgebra ${\bf g}$. As the  inner
product (8) is invariant we have $A^{-st}:{\bf {\tilde
g}}\longrightarrow {\bf g}'.i$,
\begin{equation}
X'_i=(-1)^k A_i^{\;\;k}
X_k,\hspace{10mm}\tilde{X}^{'\;j}=(A^{-st})^j_{\;\;l}\tilde{X}^l,\hspace{10mm}<X'_i
, \tilde{X}^{'\;j}> = \delta_i^{\;\;j},
\end{equation}
where $A^{-st}$ is superinverse supertranspose of every matrix
$A\in Aut(\bf g)$ [$Aut(\bf g)$ is the automorphism supergroup of
the Lie superalgebra $\bf g$]. Thus we have the following
transformation relation for the map $A^{-st}$:
\begin{equation}
(-1)^{k(j+l)} (A^{-st})^i_{\;\;k}{{\tilde{f}^{kl}}_{(\bf {\tilde
g})}}\hspace{1mm}_m (A^{-st})^j_{\;\;l} = {{{{f}}^{ij}}_{(\bf
{g}'.i)}}\hspace{0.5mm}_n (A^{-st})^n_{\;\;m}.\\
\end{equation}
Now, for obtaining Lie super-bialgebras $({\bf g}, {\bf g}'.i)$,
we must find Lie superalgebras ${\bf g}'.i$ or transformations \\
$B:{\bf g}'\longrightarrow {\bf g}'.i$, such that
\begin{equation}
(-1)^{k(j+l)} B^i_{\;\;k}{{{f}^{kl}}_{(\bf {
g}')}}\hspace{0.5mm}_m B^j_{\;\;l} = {{{{f}}^{ij}}_
{(\bf {g}'.i)}}\hspace{0.5mm}_n B^n_{\;\;m}.\\
\end{equation}
For this purpose, it is enough to omit ${{{{f}}^{ij}}_ {(\bf
{g}'.i)}}\hspace{1mm}_n $ between (19) and (20). Then we will have
the following matrix equation for $B$:
\begin{equation}
(A^{-st})^i_{\;\;m}\tilde{\cal X}^{st\;m}_{\tilde{(\bf
g)}}A^{-1} = (-1)^{i(l+j)}(B^{st} A)^{-1}(B^i_{\;\;k}{{\cal X}^{st\;k}}_{({\bf g}')}) B^{st},\\
\end{equation}
where indices $l$ and $j$ are row and column of $B^{st}$ in the
right hand side of the above relation.  Note that for $A^{-st}$
and $A^{-1}$ in $(21)$ we use row and column indices on the right
hand sides. Now by solving (21) we obtain the general form of
matrix $B$ with the condition $sdetB \neq 0$. In solving (21) one
can obtain conditions on elements of matrix $A$, yet we must only
consider those conditions under which we have $sdet A\neq 0$ and
matrices $A$, $B$ and $A^{-st}$ have the general transformation
matrix form (38).

\bigskip

{\it Step 3:}~~{\it Obtaining and classificating the
nonequivalent  Lie super-bialgebras}

\smallskip

Having solved (21), we obtain the general form of the matrix $B$
so that its elements are written in terms of the elements of
matrices $A$, $C$ and structure constants
${{\tilde{f}^{ij}}_{(\bf {\tilde g})}}\hspace{0.5mm}_k $. Now with
substituting $B$ in $(20)$, we obtain structure constants
${{{{f}}^{ij}}_ {(\bf {g}'.i)}}\hspace{0.5mm}_n $ of the Lie
superalgebra ${\bf g}'.i$ in terms of elements of matrices $A$ and
$C$ and some ${{\tilde{f}^{ij}}_{(\bf {\tilde g})}}\hspace{1mm}_k
$. Then we check whether it is possible to equalize the structure
constants ${{{{f}}^{ij}}_ {(\bf {g}'.i)}}\hspace{0.5mm}_n$ with
each other and with $\pm1$ or not so as to remark $sdet B\neq0$,
$sdet A\neq0$, and $sdet C\neq0$.
 In this way, we obtain isomorphism matrices $B_1$, $B_2$,....
 As $B_i^{st}$ must be in the form of  transformation matrices (38), we obtain conditions on the matrices
 $B_i$. The reason is
that if $({\bf g}, {\bf g}'.i)$ is Lie super-bialgebra, then $(
{\bf g}',{\bf g}.i)$ will be Lie super-bialgebras with
$B^{st}_i:{\bf g}\longrightarrow {\bf g}.i$. Note that in
obtaining $B_i$s we impose the condition $B{B_i}^{-1}\in
Aut^{st}(\bf g)$ [$Aut^{st}(\bf g)$ is the supertranspose of
$Aut(\bf g)$]; if this condition is not satisfied then we cannot
impose it on the structure constants because $B$ and $B_i$ are not
equivalent (see bellow).

\smallskip

Now using isomorphism matrices $B_1$, $B_2$, etc, we can obtain
Lie super-bialgebras $({\bf g}, {\bf g}'.i)$, $({\bf g}, {\bf
g}'.ii)$, etc. On the other hand, there is the question: which of
these Lie super-bialgebras are equivalent? In order to answer this
question, we use the matrix form of the relation (7). Consider the
two Lie super-bialgebras $({\bf g}, {\bf g}'.i)$, $({\bf g}, {\bf
g}'.ii)$; then using \cite{F3}
\begin{equation}
A(X_i)=(-1)^j A_i^{\;\;j} X_j,
\end{equation}
relation (7) will have the following matrix form:
\begin{equation}
(-1)^{i(j+l)} A^{st} ((A^{st})^i_{\;\;k}{{\cal X}_{({\bf g}'.i)}}^k) = {{{\cal X}}_{({\bf g}'.ii)}}^i A^{st}.\\
\end{equation}
On the other hand, the transformation matrix between ${\bf g}'.i$
and ${\bf g}'.ii$ is $B_2B_1^{-1}$ if $B_1:{\bf g}'\longrightarrow
{\bf g}'.i$ and $B_2:{\bf g}'\longrightarrow {\bf g}'.ii$; then we
have
\begin{equation}
(-1)^{i(j+l)} (B_2B_1^{-1}) ((B_2B_1^{-1})^i_{\;\;k}{{\cal X}_{({\bf g}'.i)}}^k) = {{{\cal X}}_{({\bf g}'.ii)}}^i (B_2B_1^{-1}).\\
\end{equation}
A comparison of $(24)$ with $(23)$ reveals that if $B_2B_1^{-1}\in
A^{st}$ holds, then the Lie super-bialgebras $({\bf g}, {\bf
g}'.i)$ and $({\bf g}, {\bf g}'.ii)$ are equivalent. In this way,
we obtain nonequivalent class of $B_i$s and we consider only one
element of this class. Thus, we obtain and classify all Lie
super-bialgebras. In the next section, we apply this formulation
to two and three dimensional Lie superalgebras.

\vspace{5mm}

\section{\bf Two and three dimensional Lie superalgebras and their automorphism supergroups }

In this section, we use the classification of two and three
dimensional Lie superalgebras listed in Ref. \cite{B}. In this
classification, Lie superalgebras are divided into two types:
trivial and nontrivial Lie superalgebras for which the
commutations of fermion-fermion is zero or nonzero, respectively
(as we use DeWitt notation here, the structure constant $C^B_{FF}$
must be pure imaginary). The results have been presented in
tables $1$ and $2$. As the tables $(m, n-m)$ indicate, the Lie
superalgebras have $m, \{X_1,...,X_m\}$ bosonic and $n-m,
\{X_{m+1},...,X_n\}$ fermionic generators.

\vspace{5mm}
\begin{center}
\hspace{10mm}{\small {\bf Table 1}} : \hspace{3mm}{\footnotesize Trivial Lie superalgebras.}\\
    \begin{tabular}{l l l l  l  p{15mm} }
    \hline\hline
   {\scriptsize Type }& {\scriptsize ${\bf g}$ }& {\scriptsize
Bosonic basis} & {\scriptsize Fermionic basis}&{\scriptsize
Non-zero commutation relations}&{\scriptsize Comments} \\ \hline
{\scriptsize $(1,1)$} & {\scriptsize $B$}& {\scriptsize $X_1$}& {\scriptsize $X_2$}&{\scriptsize $[X_1,X_2]=X_2$} \\
\hline {\scriptsize $(2,1)$} & {\scriptsize ${C^1_p}$} &
{\scriptsize $X_1,X_2$} & {\scriptsize $X_3$}&{\scriptsize
$[X_1,X_2]=X_2,\; [X_1,X_3]=pX_3 $}&{\scriptsize  $p\neq0$ } \\
\hline &{\scriptsize  $ C^2_p$} & {\scriptsize  $X_1$} &
{\scriptsize  $X_2,X_3$}& {\scriptsize
$[X_1,X_2]=X_2, \;[X_1,X_3]=pX_3$ }&{\scriptsize  $0<|p|\leq 1$} \\
 & {\scriptsize  $C^3$} & {\scriptsize  $X_1$} &
{\scriptsize  $X_2,X_3$}  &{\scriptsize  $[X_1,X_3]=X_2
$}&{\scriptsize  Nilpotent } \\
{\scriptsize  $(1,2)$} &{\scriptsize  $C^4$} & {\scriptsize
$X_1$} & {\scriptsize  $X_2,X_3$}&{\scriptsize  $[X_1,X_2]=X_2,\; [X_1,X_3]=X_2+X_3$}\\
 &{\scriptsize $C^5_p$} & {\scriptsize  $X_1$} &
{\scriptsize  $X_2,X_3$}&{\scriptsize $[X_1,X_2]=pX_2-X_3,\;
[X_1,X_3]=X_2+pX_3$ }&{\scriptsize  $p\geq0 $}
\\ \hline\hline
    \end{tabular}
\end{center}

\vspace{5mm}
\begin{center}
\hspace{10mm}{\small{\bf Table 2}} : \hspace{3mm}{\footnotesize
Nontrivial Lie superalgebras} (\cite{F4}).\\
    \begin{tabular}{l l l l  l  p{11mm} }
    \hline\hline
   {\scriptsize Type }& {\scriptsize ${\bf g}$ }& {\scriptsize
Bosonic basis} & {\scriptsize Fermionic basis}&{\scriptsize
Non-zero (anti)commutation relations}&{\scriptsize Comments} \\
\hline {\scriptsize $(1,1)$} & {\scriptsize $(A_{1,1}+A)$}&
{\scriptsize $X_1$}& {\scriptsize $X_2$} &{\scriptsize
$\{X_2,X_2\}=iX_1$}
\\\hline {\scriptsize $(2,1)$} &{\scriptsize
$C^1_{\frac{1}{2}}$ } & {\scriptsize $X_1,X_2$} & {\scriptsize
$X_3$}&{\scriptsize $[X_1,X_2]=X_2,\;
[X_1,X_3]=\frac{1}{2}X_3,\;\{X_3,X_3\}=iX_2 $ }
\\\hline
 &{\scriptsize $(A_{1,1}+2A)^1$}&
{\scriptsize $X_1$} & {\scriptsize $X_2,X_3$}&{\scriptsize
$\{X_2,X_2\}=iX_1,\; \{X_3,X_3\}=iX_1 $} &
{\scriptsize Nilpotent} \\
\vspace{-3mm} {\scriptsize $(1,2)$} &&  & & &
 \\
\vspace{-3mm} & {\scriptsize $(A_{1,1}+2A)^2$}& {\scriptsize
$X_1$} & {\scriptsize $X_2,X_3$}& {\scriptsize $\{X_2,X_2\}=iX_1,
\;\{X_3,X_3\}=-iX_1 $}&{\scriptsize  Nilpotent }
\\\hline\hline
    \end{tabular}
\end{center}

\vspace{5mm}

One can check whether super Jacobi identities for the above Lie
superalgebras are satisfied. As mentioned in section 3 for
obtaining dual Lie superalgebras we need automorphism supergroups
of Lie superalgebras. In order to calculate the automorphism
supergroups of two and three dimensional Lie superalgebras, we use
the following transformation:
\begin{equation}
{X'_i}=(-1)^j\;A_i^{\;\;j} X_j,\hspace{20mm} [X'_i ,X'_j] =f^k_{\; \; ij} X'_k, \\
\end{equation}
thus we have the following matrix equation for elements of
automorphism supergroups:
\begin{equation}
(-1)^{ij+mk} A {\cal Y}^k A^{st} = {\cal Y}^e A_e^{\;\;k},\\
\end{equation}
where index $j$ is the column of matrix $A$ and indices $i$ and
$m$ are row and column of $A^{st}$ in the left hand side. Using
(26) with the condition $sdet A\neq 0$ and imposing the condition
that $A$ must be in the form (38), we obtain table $3$ for
automorphism supergroups.\\

\vspace{5mm}

{\small {\bf Table 3} }: {\footnotesize Automorphism supergroups
of the two
and three dimensional Lie superalgebras.}\\
    \begin{tabular}{ l l  l l p{30mm} }
    \hline\hline
{\footnotesize $\bf g$} & {\footnotesize Automorphism
supergroups}&~~~~~~~{\footnotesize Comments}\\ \hline
\vspace{2mm}{\footnotesize $B$} &{\footnotesize
$\left(\begin{array}{cc}
                      1 & 0 \\
                      0 & b    \\
                     \end{array} \right)$}&~~~~~~~{\footnotesize $b \in \Re-\{0\}$} \\

\vspace{2mm}

{\footnotesize $ (A_{1,1}+A)$}  & {\footnotesize
$\left(\begin{array}{cc}
                       a^2 & 0 \\
                       0 & a \\
                       \end{array} \right)$ }&~~~~~~~{\footnotesize $a\in\Re-\{0\}$} \\
\vspace{2mm}

{\footnotesize $ C^1_p$} & {\footnotesize $\left(
\begin{array}{ccc}
                      1 & a & 0 \\
                       0 & c & 0 \\
                       0 & 0 & d
                     \end{array} \right)$}&~~~~~~~{\footnotesize $p \in \Re-\{0\},\;\;\;c,d\in\Re-\{0\}$,\; $a\in\Re$ } \\

\vspace{2mm}

{\footnotesize $ C^1_\frac{1}{2}$} &{\footnotesize
$\left(\begin{array}{ccc}
                      1 & a & 0 \\
                       0 & b^2 & 0 \\
                       0 & 0 & b
                     \end{array} \right)$}&~~~~~~~{\footnotesize $b\in\Re-\{0\}$ , $a\in\Re$ }\\

\vspace{2mm}

{\footnotesize $C^2_1$} &{\footnotesize $\left(\begin{array}{ccc}
                       1  & 0  &  0 \\
                       0 & c & b \\
                       0 & a & d
                     \end{array} \right)$}&~~~~~~~{\footnotesize $ab-
                     cd\neq 0$ }\\

\vspace{2mm}

{\footnotesize $ C^2_p$ }  & {\footnotesize
$\left(\begin{array}{ccc}
                      1 & 0 & 0 \\
                       0 & c & 0 \\
                       0 & 0 & d
                     \end{array} \right)$}&~~~~~~~{\footnotesize $c,d\in\Re-\{0\},\;\;p \in [-1,1]-\{0\}$}  \\

\vspace{2mm}

{\footnotesize $ C^3$}    &{\footnotesize $\left(
\begin{array}{ccc}
                      a & 0 & 0 \\
                       0 & ad & 0 \\
                       0 & e & d
                     \end{array} \right)$} &~~~~~~~{\footnotesize  $a,d\in\Re-\{0\}$, $e\in\Re$} \\

\vspace{2mm}

{\footnotesize $ C^4$}    & {\footnotesize $\left(
\begin{array}{ccc}
                      1 & 0 & 0 \\
                       0 & c & 0 \\
                       0 & d & c
                     \end{array} \right)$} &~~~~~~~ {\footnotesize $c\in\Re-\{0\},\;\;d\in\Re$} \\

\vspace{2mm}

{\footnotesize $ C^5_p$ }  &{\footnotesize $\left(
\begin{array}{ccc}
                      1 & 0 & 0 \\
                       0 & c & -d \\
                       0 & d & c
                     \end{array} \right)$} &~~~~~~~{\footnotesize $p\geq 0,\;\;c\in\Re-\{0\}$ or $d\in\Re-\{0\}$} \\

\vspace{2mm}

{\footnotesize $(A_{1,1}+2A)^1$ }  & {\footnotesize
$\left(\begin{array}{ccc}
                       b^2+c^2 & 0 & 0 \\
                       0 & b & -c \\
                       0 & c & b
                     \end{array} \right)$} &~~~~~~~{\footnotesize $b\in\Re-\{0\}\verb""$}
                       {\footnotesize  or} {\footnotesize $c\in\Re-\{0\}\verb""$} \\

\vspace{2mm}

{\footnotesize $ (A_{1,1}+2A)^2$}  &{\footnotesize
$\left(\begin{array}{ccc}
                      b^2-c^2 & 0 & 0 \\
                       0 & b & c \\
                       0 & c & b
                     \end{array} \right)$}&~~~~~~~ {\footnotesize $b\in\Re-\{0\}\verb""$}
                     {\footnotesize  or} {\footnotesize $c\in\Re-\{0\}\verb""$} \\\hline\hline
 \end{tabular}

\vspace{5mm}
\section{\bf Two and three dimensional Lie super-bialgebras }

Using the methods discussed in section $3$, we can classify two
and three dimensional Lie super-bialgebras. We have applied MAPLE
$9$ and obtained $48$ Lie super-bialgebras. These have been listed
in the following tables 4-6:

\vspace{8mm}

\hspace{4mm}{\small{\bf Table 4}} : \hspace{1mm}{\small Two
dimensional
Lie super-bialgebras of the type $(1 ,1)$.}\\
    \begin{tabular}{l l l l p{5mm} }
    \hline\hline
{\footnotesize ${\bf g}$ }& \hspace{1cm}{\footnotesize $\tilde{\bf
g}$
}&\hspace{2cm}{\footnotesize Non-zero (anti) commutation relations of $\tilde{\bf g}$}\\
\hline

{\footnotesize $(A_{1,1}+A)$} &\hspace{0.9cm} {\footnotesize
$I_{(1,1)}$}&
\\\hline
\vspace{1mm}

 & \hspace{1cm}{\footnotesize $I_{(1,1)}$}
\\

\vspace{1mm}

{\footnotesize $B$}&\hspace{1cm}{\footnotesize $(A_{1,1}+A)$}
&\hspace{2cm} {\footnotesize $\{{\tilde X}^2,{\tilde
X}^2\}=i{\tilde X}^1$}&
\\

\vspace{-1mm}

&\hspace{1cm}{\footnotesize $(A_{1,1}+A).i$} &
\hspace{2cm}{\footnotesize $\;\{{\tilde X}^2,{\tilde X}^2\}=-i
{\tilde X}^1$}&
\smallskip \\
\hline\hline
\end{tabular}

\vspace{5cm}

\hspace{3mm}{\small {\bf Table 5}}: \hspace{2mm}{\small
Three dimensional  Lie super-bialgebras of the type (2 ,1).}\\
    \begin{tabular}{l l l l  p{0.15mm} }
    \hline\hline
{\footnotesize ${\bf g}$ }& {\footnotesize $\tilde{\bf g}$}
&{\footnotesize Non-zero (anti) commutation relations of
$\tilde{\bf g}$}& {\footnotesize Comments} \\ \hline

\vspace{-1mm}

&{\footnotesize $I_{(2 , 1)} $}& &&\\

\vspace{-1mm}

{\footnotesize $C^1_p$}&&&\\

\vspace{2mm}

&{\footnotesize $C^1_{-p}.i$ }&{\footnotesize $[{\tilde
X}^1,{\tilde X}^2]={\tilde X}^1,\;\;\;[{\tilde X}^2,{\tilde
X}^3]=p{\tilde X}^3$} &{\footnotesize $p\in\Re-\{0\}$}
\\\hline

\vspace{2mm}

&{\footnotesize $I_{(2 , 1)} $}&\\

\vspace{1mm}

&{\footnotesize
$C^1_{p}.i{_{|_{p=-\frac{1}{2}}}}$}&{\footnotesize$[{\tilde
X}^1,{\tilde X}^2]= {\tilde X}^1,\;\;\;[{\tilde X}^2,{\tilde
X}^3]=\frac{1}{2} {\tilde X}^3$}&
 \\
\vspace{-2mm}

&{\footnotesize
$C^1_{p}.ii{_{|_{p=-\frac{1}{2}}}}$}&{\footnotesize$[{\tilde
X}^1,{\tilde X}^2]=-{\tilde X}^1,\;\;\;[{\tilde X}^2,{\tilde
X}^3]=-\frac{1}{2} {\tilde X}^3$}&
 \\
\vspace{-2mm}

{\footnotesize $C^1_{\frac{1}{2}}$}& \\

\vspace{2mm}

&{\footnotesize $C^1_{\frac{1}{2}}.i$ }&{\footnotesize $[{\tilde
X}^1,{\tilde X}^2]={\tilde X}^1,\;\;\;[{\tilde X}^2,{\tilde
X}^3]=-\frac{1}{2} {\tilde X}^3,\;\;\; \{{\tilde X}^3,{\tilde
X}^3\}=i{\tilde X}^1$}&
\\

\vspace{2mm}

&{\footnotesize $C^1_{\frac{1}{2}}.ii$ }&{\footnotesize $[{\tilde
X}^1,{\tilde X}^2]=-{\tilde X}^1,\;\;\;[{\tilde X}^2,{\tilde
X}^3]=\frac{1}{2} {\tilde X}^3,\;\;\; \{{\tilde X}^3,{\tilde
X}^3\}=-i{\tilde X}^1$}&
\\

\vspace{1mm}

&{\footnotesize ${{C^1_{\frac{1}{2},k}}}$}&{\footnotesize
$[{\tilde X}^1,{\tilde X}^2]=k{\tilde X}^2,\;[{\tilde X}^1,{\tilde
X}^3]=\frac{k}{2}{\tilde X}^3,\;\{{\tilde X}^3,{\tilde
X}^3\}=ik{\tilde X}^2$}&{\footnotesize $k\in {\Re-\{0\}}$}
\smallskip \\
\hline\hline

 \end{tabular}

 \vspace{10mm}


\hspace{2mm}{\small {\bf Table 6}}: \hspace{1mm}{\small Three
dimensional  Lie super-bialgebras of the type (1 ,2). Where
$\epsilon=\pm 1$.}\\
  \begin{tabular}{l l l  p{0.5mm} }
    \hline\hline
 {\footnotesize ${\bf g}$ }& {\footnotesize  $\tilde{\bf g}$}
  &{\footnotesize  Comments }\\ \hline
\vspace{2mm}
 &{\footnotesize $I_{(1 , 2)} $}&\\
\vspace{2mm}

{\footnotesize $C^2_1$}&{\footnotesize $(A_{1,1}+2A)^1_{\epsilon,0,\epsilon}$ } & & \\
\vspace{2mm}

&{\footnotesize  $(A_{1,1}+2A)^2_{\epsilon,0,-\epsilon}$ }& &
\\\hline

\vspace{2mm}
&{\footnotesize  $I_{(1 , 2)} $}&\\
\vspace{2mm}

{\footnotesize $C^2_p$}& {\footnotesize $(A_{1,1}+2A)^1_{\epsilon
, k,\epsilon }$}
 &{\footnotesize $ -1<k<1$}& \\
\vspace{2mm}

{\footnotesize $p \in [-1,1)-\{0\}$ }&{\footnotesize
$(A_{1,1}+2A)^2_{0
,1,0}\;,\;\;(A_{1,1}+2A)^2_{\epsilon,1,0}\;,\;\;(A_{1,1}+2A)^2_{0,1,\epsilon}\;,\;\;
(A_{1,1}+2A)^2_{\epsilon ,k,-\epsilon }$} &{\footnotesize
$k\in\Re$}& \\\hline

\vspace{2mm}

& {\footnotesize $I_{(1 , 2)} $}&\\
\vspace{2mm}

{\footnotesize $C^3$}&{\footnotesize  $(A_{1,1}+2A)^1_{\epsilon ,
0,\epsilon }$}
 && \\

\vspace{2mm}

&{\footnotesize  $(A_{1,1}+2A)^2_{0 ,\epsilon,0
}\;,\;\;(A_{1,1}+2A)^2_{\epsilon ,0,-\epsilon}$}
 && \\\hline

\vspace{2mm}
&{\footnotesize  $I_{(1 , 2)} $}&\\
\vspace{2mm}

{\footnotesize $C^4$}&{\footnotesize  $(A_{1,1}+2A)^1_{k ,
0,1}\;,\;\;(A_{1,1}+2A)^1_{s , 0,-1}$}
 &{\footnotesize $0<k,\;\;s<0$}& \\

\vspace{2mm} &{\footnotesize  $(A_{1,1}+2A)^2_{0 ,\epsilon,0
}\;,\;\;(A_{1,1}+2A)^2_{k ,0,1}\;,\;\;(A_{1,1}+2A)^2_{s,0,-1}$}
 &{\footnotesize $k<0,\;\;0<s$}\\\hline

\vspace{2mm}
& {\footnotesize $I_{(1 , 2)} $}&\\
\vspace{2mm}

{\footnotesize $C^5_p$}&{\footnotesize  $(A_{1,1}+2A)^1_{k ,
0,1}\;,\;\;(A_{1,1}+2A)^1_{s , 0,-1}$}
 &{\footnotesize $0<k,\;\;s<0$}& \\

{\footnotesize $p\geq 0$}&{\footnotesize  $(A_{1,1}+2A)^2_{k
,0,1}\;,\;\;(A_{1,1}+2A)^2_{s,0,-1}$}
 &{\footnotesize $ k<0,\;\;0<s$}\smallskip\smallskip\\\hline

\vspace{2mm}

{\footnotesize $(A_{1,1}+2A)^1$}&{\footnotesize  $I_{(1 , 2)} $}&\\

\vspace{1mm}

{\footnotesize $(A_{1,1}+2A)^2$}& {\footnotesize $I_{(1 , 2)} $}& \smallskip \\
\hline\hline
 \end{tabular}

\vspace{1mm}

\vspace{4mm}

For three dimensional dual Lie superalgebras
$(A_{1,1}+2A)^1_{\alpha ,\beta,\gamma }\;$ and $\;
(A_{1,1}+2A)^2_{\alpha ,\beta,\gamma }$ which are isomorphic with
$(A_{1,1}+2A)^1$ and  $(A_{1,1}+2A)^2$, respectively, we have the
following commutation relations:

\begin{equation}
\{{\tilde X}^2,{\tilde X}^2\}=i\alpha {\tilde X}^1 ,\quad
\{{\tilde X}^2,{\tilde X}^3\}=i \beta {\tilde X}^1 ,\quad
\{{\tilde X}^3,{\tilde X}^3\}=i\gamma {\tilde X}^1,\qquad \alpha,
\beta, \gamma \in\Re.
\end{equation}
Note that these Lie superalgebras are non isomorphic and they
differ in the bound of their parameters.

\smallskip

Meanwhile note that for every Lie superalgebra of tables 1 and 2
there are trivial solutions for Eqs. (14) and (15) with ${\tilde
f}^{ij}_{\;\;k}=0$. (i.e. the dual Lie superalgebra is Abelian)
we denote these dual Lie superalgebras with $I_{(m,n)}$.

We see that for two and three dimensional Lie super-bialgebras
(with two fermions), trivial Lie superalgebras are only dual to
nontrivial one. The solution of super Jacobi and mixed super
Jacobi identities and related isomorphism matrices $C$ (in step 1)
are listed in Appendix $B$. Here for explanation of steps 1-3 of
section 3 we give an example.

\vspace{5mm}
\subsection{\bf An example}
\vspace{4mm}

What follows is an explanation of the details of calculations for
obtaining dual Lie superalgebras of Lie super-bialgebras $(C^4 ,
(A_{1,1}+2A)^1_{k , 0,1})$ and $(C^4 , (A_{1,1}+2A)^1_{s ,
0,-1})$. As mentioned in Appendix $B$, the solution of super
Jacobi and mixed super Jacobi identities has the following form:
\begin{equation}
{{\tilde{f}}^{22}}_{\; \; \: 1}=i\alpha,\qquad
{{\tilde{f}}^{33}}_{\; \; \; 1}=i\beta ,\qquad
{{\tilde{f}}^{23}}_{\; \; \; 1}=i\gamma,\;\;\;\;\forall \alpha,
\beta, \gamma \in \Re. \hspace{3cm}
\end{equation}
This solution is isomorphic with the Lie superalgebra
$(A_{1,1}+2A)^1$ with the following matrix:
\begin{equation}
C_1=\left( \begin{tabular}{ccc}
              $ c_{11}$&  0 & 0 \\
                $ c_{21}$ & $c_{22}$ & $ c_{23}$ \\
                 $c_{31}$ & 0 & $ c_{33}$ \\
                \end{tabular} \right), \hspace{2cm}
\end{equation}
where  $c_{11}= -ic^2_{33} {{\tilde{f}}^{33}}_{\; \; \; 1},\;
c_{23}= -c_{22} \frac{{{\tilde{f}}^{23}}_{\; \; \;
1}}{{{\tilde{f}}^{33}}_{\; \; \; 1}},\; {{\tilde{f}}^{22}}_{\; \;
\;1}=(\frac{c^2_{23}+c^2_{33}}{c^2_{22}}) {{\tilde{f}}^{33}}_{\;
\; \; 1}$ and $c_{22}$, $c_{33}\in\Re-\{0\}$. By imposing that
$C_1$ must be the transformation matrix $(38)$, we have
$c_{31}=c_{21}=0$.\\
Now with the help of automorphism supergroup of $C^4$, the
solution of $(21)$ for the matrix $B$ will be
\begin{equation}
B=\left( \begin{tabular}{ccc}
              $ \frac{c^2(b^2_{32}+b^2_{33})}{\beta}$&  0 & 0 \\
                $ 0$ & $-\frac{b_{33}c_{33}}{c_{22}}$ & $ \frac{b_{32}c_{33}}{c_{22}}$\\
                 $0$ & $b_{32}$ & $b_{33} $ \\
                \end{tabular} \right),\;\;\;\beta=-i {{\tilde{f}}^{33}}_{\; \; \; 1}.\hspace{2cm}
\end{equation}
Now using $(20)$, we obtain the following commutation relations
for the dual Lie superalgebra ${(A_{1,1}+2A)^1}{'}$:
\begin{equation}
\{{\tilde X}^2,{\tilde X}^2\}=ia^{'}{\tilde X}^1,\qquad \{{\tilde
X}^3,{\tilde X}^3\}=ib^{'}{\tilde
X}^1,\;\;\;\;\;\;a^{'}=\frac{\beta c^2_{33}}{c^2
c^2_{22}},\;\;b^{'}=\frac{\beta}{c^2},
\end{equation}
such that $a^{'}$ and $b^{'}$ have the same
sings.\\
One cannot  choose $a^{'}=b^{'}$ because in this case we have
\begin{equation}
B_1=\left( \begin{tabular}{ccc}
              $\frac{{c}^2({b^{'}}^2_{32}+{b^{'}}^2_{33})}{\beta}$&  0 & 0 \\
                $ 0$ & $-b^{'}_{33}$ & $b^{'}_{32}$ \\
                 $0$ &$b^{'}_{32}$ & $b^{'}_{33}$ \\
                \end{tabular} \right), \hspace{2cm}
\end{equation}
so as $BB^{-1}_1\notin A^{st}(C^4)$. Now by choosing $b^{'}=1$
i.e., $\beta=c^2$ we have
\begin{equation}
B_2=\left( \begin{tabular}{ccc}
              ${b^{''}}^2_{32}+{b^{''}}^2_{33}$&  0 & 0 \\
                $ 0$ & $-\frac{c^{''}_{33}}{c^{''}_{22}}b^{''}_{33}$ & $\frac{c^{''}_{33}}{c^{''}_{22}}b^{''}_{32}$ \\
                 $0$ & $b^{''}_{32}$ & $b^{''}_{33} $ \\
                \end{tabular} \right). \hspace{2cm}
\end{equation}
such that $BB^{-1}_2\in A^{st}(C^4)$, this means that we can choose $b^{'}=1$ and in
this case, we have the following dual Lie superalgebras ($a^{'}=\frac{c^2_{33}}{c^2_{22}}=k>0$): \\
\begin{equation}
(A_{1,1}+2A)^1_{k , 0,1}:\hspace{1cm}   \quad  \{{\tilde
X}^2,{\tilde X}^2\}=ik{\tilde X}^1,\qquad \{{\tilde X}^3,{\tilde
X}^3\}=i{\tilde
 X}^1,\;\;k>0. \hspace{1cm}
\end{equation}
In the same way by choosing $b^{'}=-1$ i.e., $\beta=-c^2$ such
that $a^{'}\neq b^{'}$ we have
\begin{equation}
B_3=\left( \begin{tabular}{ccc}
              $-({b^{'''}}^2_{32}+{b^{'''}}^2_{33})$&  0 & 0 \\
                $ 0$ & $-\frac{c^{'''}_{33}}{c^{'''}_{22}}b^{'''}_{33}$ & $\frac{c^{'''}_{33}}{c^{'''}_{22}}b^{'''}_{32}$ \\
                 $0$ & $b^{'''}_{32}$ & $b^{'''}_{33} $ \\
                \end{tabular} \right). \hspace{2cm}
\end{equation}
so as $BB^{-1}_3\in A^{st}(C^4)$, this means that we can choose
$b^{'}=-1$ and in
this case, we have the following dual Lie superalgebras ($a^{'}=-\frac{c^2_{33}}{c^2_{22}}=s<0$): \\
\begin{equation}
(A_{1,1}+2A)^1_{s , 0,-1}:\hspace{1cm} \{{\tilde X}^2,{\tilde
 X}^2\}=is{\tilde X}^1,\qquad \{{\tilde
 X}^3,{\tilde X}^3\}=-i{\tilde X}^1,\;\;s<0. \hspace{1cm}
\end{equation}
Note that as $B_2B^{-1}_3\notin A^{st}(C^4)$, this means that the
Lie super-bialgebras $(C^4 , (A_{1,1}+2A)^1_{k , 0,1})$ and $(C^4
, (A_{1,1}+2A)^1_{s , 0,-1})$ are not equivalent.

\smallskip

\section{\bf Conclusion}
We have presented a new method for obtaining and classifying low
dimensional {\it Lie super-bialgebras} which can also be applied
for classification of low dimensional {\it Lie bialgebras}. We
have classified all of the $48$ two and three dimensional Lie
super-bialgebras \cite{F5}. Using this classification one can
construct Poisson-Lie T-dual sigma models \cite{K.S1} on the low
dimensional Lie supergroups \cite{ER} (as Ref. \cite{JR} for Lie
groups). Moreover, one can determinate  the coboundary type of
these Lie super-bialgebras and their classical r-matrices and
Poisson brackets for their Poisson-Lie supergroups \cite{ER1}.
Determination of doubles and their isomorphism and investigation
of Poisson-Lie plurality \cite{Von} are open problems for further
investigation.\\
\vspace{5mm}

{\bf Acknowledgments}

\vspace{3mm} We would like to thank S. Moghadassi and F. Darabi
for carefully reading the manuscript and useful comments.

\vspace{8mm}

{\bf Appendix A: Some properties of matrices and tensors in
supervector spaces}

\vspace{4mm}

In this appendix, we will review some basic
properties of tensors and matrices in Ref. \cite{D}.\\

\hspace{0.55cm}1. We consider the standard basis for the
supervector spaces so that in writing the basis as a column
matrix, we first present the bosonic base, then the fermionic
one. The transformation of standard basis and its dual bases can
be written as follows:
\begin{equation}
 {e'}_i=(-1)^j{K_i}\;^j e_j ,\hspace{10mm}{e'}^i={{K^{-st}}^i}_j\; e^j,
\end{equation} where the transformation matrix $K$ has the
following block diagonal representation \cite{D}
\begin{equation}
K=\left( \begin{tabular}{c|c}
                 A & C \\ \hline
                 D & B \\
                 \end{tabular} \right),
\end{equation} where $A,B$ and $C$ are real submatrices and $D$ is pure
imaginary submatrix \cite{D}. Here we consider the matrix and
tensors having a form with all upper and lower indices written in
the right hand side.

\hspace{0.55cm}2. The transformation properties of upper and lower
right indices to the left one for general tensors are as follows:
\begin{equation}
^iT_{jl...}^{\;k}=T_{jl...}^{ik},\qquad
_jT^{ik}_{l...}=(-1)^j\;T_{jl...}^{ik}.
\end{equation}
\hspace{1cm}3. For supertransposition we have
$$
{{L}^{st\;i}}_j=(-1)^{ij}\;{L_j}^{\;i},\qquad
{L^{st}_{\;i}}^{\;j}=(-1)^{ij}\;{L^i}_{\;j},
$$
\begin{equation}
M^{st}_{\;ij}=(-1)^{ij}\;M_{\;ji},\qquad
M^{st\;ij}=(-1)^{ij}\;M^{\;ji}.
\end{equation}
\hspace{1cm}4. For superdeterminant we have
\begin{equation}
sdet\left( \begin{tabular}{c|c}
                 A & C \\ \hline
                 D & B \\
                 \end{tabular} \right)=det{(A-CB^{-1}D)}(det B)^{-1},
\end{equation}
when $det B\neq0$ and
\begin{equation}
sdet\left( \begin{tabular}{c|c}
                 A & C \\ \hline
                 D & B \\
                 \end{tabular} \right)=(det{(B-DA^{-1}C)})^{-1}\;(det A),
\end{equation}
when $det A\neq0$. For the inverse of matrix we have
\begin{equation}
{\footnotesize \left( \begin{tabular}{c|c}
                 A & C \\ \hline
                 D & B \\
                 \end{tabular} \right)^{-1}=\left( \begin{tabular}{c|c}
                 $(1_m-A^{-1}C
                  B^{-1}D)^{-1}A^{-1}$&
                  $-(1_m-A^{-1}CB^{-1}D)^{-1}A^{-1}CB^{-1}$  \\ \hline

                 $-(1_n-B^{-1}DA^{-1}C)^{-1}B^{-1}DA^{-1}$  & $(1_n-B^{-1}DA^{-1}C)^{-1}B^{-1}$
                \end{tabular} \right),}
\end{equation}
where $det A,det B\neq0$ and $m$,$n$ are dimensions of
submatrices $A$ and $B$, respectively.

\vspace{8mm}

{\bf Appendix B: Solutions $(sJ-msJ)$ for dual Lie superalgebras
and isomorphism matrices}
 \vspace{5mm}

This appendix includes solutions of super Jacobi and mixed super
Jacobi identities $(sJ-msJ)$ for dual Lie superalgebras and
isomorphism matrices $C$ which relate these solutions to other Lie
superalgebras.

1. Solution of $(sJ-msJ)$ for dual Lie superalgebras of $B$ is
\begin{equation}
{{\tilde{f}}^{22}}_{\;
\;\:1}=i\alpha,\qquad{{\tilde{f}}^{12}}_{\; \; \: 1}=0,
\end{equation}
where $\alpha\in\Re$.\\
Isomorphism matrix between these solutions and $(A_{1,1}+A)$ is as
follows:
\begin{equation}
C=\left( \begin{tabular}{cc}
              $ -ic^2_{22} {{\tilde{f}}^{22}}_{\; \; \: 1} $ & 0\\
                $ c_{21}$ & $c_{22}$ \\
                 \end{tabular} \right), \hspace{2cm}
\end{equation}
with the conditions $c_{22}\in\Re-\{0\}$ and imposing that $C$
must be the transformation matrix, we have $c_{21}=0$.

2. Solutions of $(sJ-msJ)$ for  dual Lie superalgebras of
$C^1_p\;(p\in\Re-\{0\})$ are

\vspace{3mm}

\begin{tabular}{l l l l p{2mm} }

\vspace{2mm}

$i)$&${{\tilde{f}}^{12}}_{\; \; \:
1}=\alpha,\;\;{{\tilde{f}}^{23}}_{\; \; \:
3}=p\alpha,$&$\;\;\;\;\;\;\;\;p\in\Re-\{0\},\;\;\;\alpha \in\Re,$&\\

\vspace{2mm}

$ii)$&${{\tilde{f}}^{12}}_{\; \; \: 1}={{\tilde{f}}^{23}}_{\; \;
\: 3}=\beta,$&$\;\;\;\;\;\;\;\;p=1,\;\;\;\beta \in\Re,$&\\

\vspace{2mm}

$iii)$ &${{\tilde{f}}^{33}}_{\; \; \:
1}=i\gamma,$ &$\;\;\;\;\;\;\;\;p\in\Re-\{0\},\;\;\;\gamma \in\Re,$&\\

\vspace{2mm}

$iv)$ &${{\tilde{f}}^{33}}_{\; \; \:
1}=i\lambda,\;\;{{\tilde{f}}^{33}}_{\; \; \:
2}=i\eta$ &$\;\;\;\;\;\;\;\;p=\frac{1}{2},\;\;\;\lambda, \eta \in\Re,$&\\

\vspace{2mm}

$v)$ &${{\tilde{f}}^{12}}_{\; \; \:
1}=\mu,\;\;{{\tilde{f}}^{23}}_{\; \; \:
3}=-\frac{\mu}{2},\;\;{{\tilde{f}}^{33}}_{\; \; \:
1}=i\nu$ &$\;\;\;\;\;\;\;\;p=-\frac{1}{2},\;\;\;\mu, \nu \in\Re.$&\\
\end {tabular}

For solution $(i)$ we have isomorphism matrix between these
solutions and ${C^1_p}$ as follows:
\begin{equation}
C=\left( \begin{tabular}{ccc}
              $ c_{11}$ & $-\frac{1}{{{\tilde{f}}^{12}}_{\; \; \: 1}}$& $ c_{13}$ \\
                $ c_{21}$ & 0 & 0 \\
                 0  & 0 & $ c_{33}$ \\
                 \end{tabular} \right), \hspace{2cm}
\end{equation}
with the conditions ${{\tilde{f}}^{23}}_{\; \; \:3}=p
{{\tilde{f}}^{12}}_{\; \; \: 1}$ and $c_{21}$,
$c_{33}\in\Re-\{0\}$, $c_{11}\in\Re$, as well as imposing that $C$
must be transformation matrix we have $c_{13}=0$. Solution $(ii)$
is a special case of solution $(i)$. For solution $(iii)$ and
$(iv)$ $sdet C=0$. Solution $(v)$ is investigated for
$(C^1_\frac{1}{2} , C^1_p)$ in
the following step.\\

3. Solutions of $(sJ-msJ)$ for dual Lie superalgebras of
$C^1_\frac{1}{2}$ are

\vspace{3mm}
\begin{tabular}{l l l l p{2mm} }

\vspace{2mm}

$i)$&${{\tilde{f}}^{23}}_{\; \; \:
3}=\alpha,\;\;\;{{\tilde{f}}^{12}}_{\; \; \:
1}=2\alpha,\;\;\;\;\alpha \in\Re,$&&\\

\vspace{2mm}

$ii)$&${{\tilde{f}}^{33}}_{\; \; \:
1}=i\beta,\;\;\;{{\tilde{f}}^{33}}_{\; \; \:
2}=i\gamma,\;\;\;{{\tilde{f}}^{23}}_{\; \; \:
3}=\frac{\beta}{2},\;\;\;{{\tilde{f}}^{13}}_{\; \; \:
3}=\frac{-\gamma}{2},\;\;\;{{\tilde{f}}^{12}}_{\; \; \:
1}=-\beta,\;\;\;{{\tilde{f}}^{12}}_{\; \; \:
2}=-\gamma,\;$&$\beta, \gamma \in\Re.$&\\

\end{tabular}

For solution $(i)$ we have isomorphism matrix $C$ between
$(C^1_\frac{1}{2} , C^1_p)$ as follows:
\begin{equation}
C=\left(\begin {tabular}{ccc}
              $ c_{11}$ & $-\frac{1}{2{{\tilde{f}}^{23}}_{\; \; \; 3}}$& $ c_{13}$ \\
                $ c_{21}$ & 0 & 0 \\
                 0  & 0 & $ c_{33}$ \\
                 \end{tabular} \right), \hspace{2cm}
\end{equation}
with the conditions ${{\tilde{f}}^{12}}_{\; \; \;3}=2
{{\tilde{f}}^{23}}_{\; \; \; 3}$, $c_{21}$, $c_{33}\in\Re-\{0\}$,
$c_{11}\in\Re$ and $p=-\frac{1}{2}$, as well as imposing that $C$ must be the transformation matrix, we have $c_{13}=0$.\\

For solution $(ii)$ we have isomorphism matrices $C_1$ and $C_2$
between $(C^1_\frac{1}{2} , C^1_\frac{1}{2})$ as follows:
\begin{equation}
C_1=\left( \begin{tabular}{ccc}
              $ c_{11}$ & $\frac{i}{{{\tilde{f}}^{33}}_{\; \; \; 1}}$& $ 0 $ \\
                $  -ic^2_{33} {{\tilde{f}}^{33}}_{\; \; \; 1}$ & 0 & 0 \\
                 0  & 0 & $ c_{33}$ \\
               \end{tabular} \right), \hspace{2cm}
\end{equation}
with the conditions ${{\tilde{f}}^{33}}_{\; \; \:2}=
{{\tilde{f}}^{13}}_{\; \; \; 3}={{\tilde{f}}^{12}}_{\; \; \;
2}=0$, $c_{33}\in\Re-\{0\}$ and $c_{11}\in\Re$\\
\begin{equation}
C_2=\left( \begin{tabular}{ccc}
              $ c_{12}\frac{{{\tilde{f}}^{33}}_{\; \; \;1}}{{{\tilde{f}}^{33}}_{\; \; \;2}}
              -\frac{i}{{{\tilde{f}}^{33}}_{\; \; \; 2}}$ & $c_{12}$& $ 0 $ \\
                $ -ic^2_{33} {{\tilde{f}}^{33}}_{\; \; \; 1}$ &$-ic^2_{33} {{\tilde{f}}^{33}}_{\; \; \; 2}$& 0 \\
                 0  & 0 & $ c_{33}$ \\
               \end{tabular} \right), \hspace{2cm}
\end{equation}
with the conditions $c_{33}\in\Re-\{0\}$ and $c_{12}\in\Re$.\\

4. Solution of $(sJ-msJ)$ for dual Lie superalgebras of $C^2_p$,
$C^3$, $C^4$ and $C^5_p$ is
\begin{equation}
{{\tilde{f}}^{22}}_{\; \; \: 1}=i\alpha,\qquad
{{\tilde{f}}^{33}}_{\; \; \; 1}=i\beta ,\qquad
{{\tilde{f}}^{23}}_{\; \; \; 1}=i\gamma, \hspace{4cm}
\end{equation} where $\alpha$,$\beta$ and $\gamma \in\Re$.
Furthermore for $C^3$ we have another solution as follows:
$$
{{\tilde{f}}^{22}}_{\; \; \: 1}=i \alpha,\qquad \alpha \in\Re,
$$
where for this solution we have $sdet C=0$, therefore we omit it. \\

\hspace{0.25cm}4.1. For the above solution, we have the following
isomorphism matrices $C_1$, $C_2$ and $C_3$ between $C^2_p$,
$C^3$, $C^4$ and $C^5_p$ with $(A_{1,1}+2A)^1$:
\begin{equation}
C_1=\left( \begin{tabular}{ccc}
              $ -ic^2_{33} {{\tilde{f}}^{33}}_{\; \; \; 1}$&  0 & 0 \\
                $ c_{21}$ & $c_{22}$ & $ -c_{22}\frac{{{\tilde{f}}^{23}}_{\; \; \;1}}{{{\tilde{f}}^{33}}_{\; \; \;1}}$ \\
                 $c_{31}$ & 0 & $ c_{33}$ \\
                \end{tabular} \right), \hspace{2cm}
\end{equation}
with the conditions ${{\tilde{f}}^{22}}_{\; \;
\;1}=(\frac{c^2_{23}+c^2_{33}}{c^2_{22}}) {{\tilde{f}}^{33}}_{\;
\; \; 1}$ and $c_{22}$, $c_{33}\in\Re-\{0\}$, as well as imposing
that $C_1$ must be the transformation matrix, we have
$c_{31}=c_{21}=0$,\\
\begin{equation}
C_2=\left( \begin{tabular}{ccc}
              $ -ic^2_{23} {{\tilde{f}}^{33}}_{\; \; \; 1}$&  0 & 0 \\
                $ c_{21}$ & $0$ & $c_{23} $ \\
                 $c_{31}$ & $-c_{33}\frac{{{\tilde{f}}^{33}}_{\; \; \;1}}{{{\tilde{f}}^{23}}_{\; \; \;1}}$& $ c_{33}$ \\
                \end{tabular} \right), \hspace{2cm}
\end{equation}
with the conditions ${{\tilde{f}}^{22}}_{\; \;
\;1}=(\frac{c^2_{23}+c^2_{33}}{c^2_{32}}) {{\tilde{f}}^{33}}_{\;
\; \; 1}$ and $c_{23}$, $c_{33}\in\Re-\{0\}$, as well as imposing
that $C_2$ must be transformation matrix, we have
$c_{31}=c_{21}=0$,\\
\begin{equation}
C_3=\left( \begin{tabular}{ccc}
              $c_{11}$&  0 & 0 \\
                $ c_{21}$ & $c_{22}$ & $c_{23} $ \\
                 $c_{31}$ & $c_{32}$& $ c_{33}$ \\
                \end{tabular} \right), \hspace{2cm}
\end{equation}
with the conditions ${{\tilde{f}}^{22}}_{\; \;
\;1}=(\frac{c^2_{23}+c^2_{33}}{c^2_{22}+c^2_{32}})
{{\tilde{f}}^{33}}_{\; \; \; 1}$, ${{\tilde{f}}^{23}}_{\; \;
\;1}=-(\frac{c_{33}c_{32}+c_{22}c_{23}}{c^2_{22}+c^2_{32}}){{\tilde{f}}^{33}}_{\;
\; \; 1}$ and $c_{11}\in\Re-\{0\},
\;\;c_{22},c_{23},c_{32},$\\$c_{33}\in\Re$ with the conditions
$c^2_{23}+c^2_{33}\neq 0$, $c^2_{22}+c^2_{32}\neq 0$, as well as
imposing that $C_3$ must be the transformation matrix, we have
$c_{31}=c_{21}=0$.\\

\hspace{0.25cm}4.2. For the above solution we have isomorphism
matrices $C_4$, $C_5$, $C_6$, $C_7$ and $C_8$ between $C^2_p$,
$C^3$, $C^4$ and $C^5_p$ with $(A_{1,1}+2A)^2$ as follows:
\begin{equation}
C_4=\left( \begin{tabular}{ccc}
              $ -ic_{32}(c_{23}-c_{33} ) {{\tilde{f}}^{23}}_{\; \; \; 1}$&  0 &0\\
                $ c_{21}$ & $c_{32}$ & $  c_{23}$ \\
                 $c_{31}$ & $c_{32}$ & $ c_{33} $ \\
                \end{tabular} \right), \hspace{2cm}
\end{equation}
with the conditions ${{\tilde{f}}^{22}}_{\;\;\;1}
=-(\frac{c_{23}+c_{33}}{c_{32}}) {{\tilde{f}}^{23}}_{\; \; \;
1}$, ${{\tilde{f}}^{33}}_{\; \; \; 1}=0$, $c_{23} \neq c_{33}$,
$c_{32}\in\Re-\{0\}$ and $c_{23},c_{33}\in\Re$, as well as
imposing that $C_4$
must be the transformation matrix, we have $c_{31}=c_{21}=0$,\\
\begin{equation}
C_5=\left( \begin{tabular}{ccc}
              $ ic_{32}(c_{23}+c_{33} ) {{\tilde{f}}^{23}}_{\; \; \; 1}$&  0 &0\\
                $ c_{21}$ & $-c_{32}$ & $  c_{23}$ \\
                 $c_{31}$ & $c_{32}$ & $ c_{33} $ \\
                \end{tabular} \right), \hspace{2cm}
\end{equation}
with the conditions ${{\tilde{f}}^{22}}_{\;\;\;1}
=(\frac{c_{23}-c_{33}}{c_{32}}) {{\tilde{f}}^{23}}_{\; \; \; 1}$,
${{\tilde{f}}^{33}}_{\; \; \; 1}=0$, $c_{23} \neq -c_{33}$,
$c_{32}\in\Re-\{0\}$ and $c_{23},c_{33}\in\Re$, as well as
imposing that $C_5$
must be the transformation matrix, we have $c_{31}=c_{21}=0$,\\
\begin{equation}
C_6=\left( \begin{tabular}{ccc}
              $ -ic^2_{23}{{\tilde{f}}^{33}}_{\; \; \; 1}$&  0 &0\\
                $ c_{21}$ & $0$ & $ c_{23}$ \\
                 $c_{31}$ & $-c_{33}\frac{{{\tilde{f}}^{33}}_{\; \; \; 1}}
                 {{{\tilde{f}}^{23}}_{\; \; \; 1}}$ & $ c_{33} $ \\
                \end{tabular} \right), \hspace{2cm}
\end{equation} with the conditions ${{\tilde{f}}^{22}}_{\;\;\;1}
=-(\frac{c^2_{23}-c^2_{33}}{c^2_{32}}) {{\tilde{f}}^{33}}_{\; \;
\; 1}$ and $c_{23}$,
$c_{33}\in\Re-\{0\}$, as well as imposing that $C_6$ must be the transformation matrix, we have $c_{31}=c_{21}=0$,\\
\begin{equation}
C_7=\left( \begin{tabular}{ccc}
              $ ic^2_{33}{{\tilde{f}}^{33}}_{\; \; \; 1}$&  0 &0\\
                $ c_{21}$ & $c_{22}$ & $ -c_{22}\frac{{{\tilde{f}}^{23}}_{\; \; \; 1}}
                 {{{\tilde{f}}^{33}}_{\; \; \; 1}}$ \\
                 $c_{31}$ & $0$ & $ c_{33} $ \\
                \end{tabular} \right), \hspace{2cm}
\end{equation}
with the conditions ${{\tilde{f}}^{22}}_{\;\;\;1}
=(\frac{c^2_{23}-c^2_{33}}{c^2_{22}}) {{\tilde{f}}^{33}}_{\; \; \;
1}$ and $c_{22}$,
$c_{33}\in\Re-\{0\}$, as well as imposing that $C_7$ must be the transformation matrix, we have $c_{31}=c_{21}=0$,\\
\begin{equation}
C_8=\left( \begin{tabular}{ccc}
              $c_{11}$&  0 & 0 \\
                $ c_{21}$ & $c_{22}$ & $c_{23} $ \\
                 $c_{31}$ & $c_{32}$& $ c_{33}$ \\
                \end{tabular} \right), \hspace{2cm}
\end{equation}
with the conditions ${{\tilde{f}}^{22}}_{\; \;
\;1}=(\frac{c^2_{33}-c^2_{23}}{c^2_{32}-c^2_{22}})
{{\tilde{f}}^{33}}_{\; \; \; 1}$, ${{\tilde{f}}^{23}}_{\; \;
\;1}=(\frac{c_{22}c_{23}-c_{32}c_{33}}{c^2_{32}-c^2_{22}}){{\tilde{f}}^{33}}_{\;
\; \; 1}$ and $c_{11}\in\Re-\{0\},
\;\;c_{22},c_{23},c_{32},$\\$c_{33}\in\Re$ with the conditions
$c^2_{33}-c^2_{23}\neq 0, c^2_{32}-c^2_{22}\neq 0$, as well as
imposing that $C_8$ must be the transformation matrix, we have
$c_{31}=c_{21}=0$.\\


\begin{thebibliography}{99}

\bibitem{Drin} V. G. Drinfel'd, \textit{ Hamiltonian Lie groups, Lie bialgebras and the
geometric meaning of the classical Yang-Baxter equation}, Sov.
Math. Dokl. \textbf{ 27} (1983) 68-71.

\bibitem{Etin} P. Etingof, D. Kazhdan, \textit{ Quantization of Lie bialgebras,} I, Selecta Math. \textbf{ 2} (1996) no. 1,
1-41.\\
P. Etingof, D. Kazhdan, \textit{ Quantization of Lie bialgebras},
II. Selecta Math. \textbf{ 4} (1998) no. 2, 213-231.

\bibitem{Del} P. Delorme, \textit{ Sur les triples de Manin pour les complexes,} Jour.
Alg. \textbf{ 246} (2001) 97-174.

\bibitem{JR}  M.A. Jafarizadeh and A. Rezaei-Aghdam, \textit{ Poisson-Lie
T-duality and Bianchi type algebras}, \textit{Phys. Lett. B}
\textbf{ 458} (1999) 477-490, \texttt{[arXiv:hep-th/9903152]}.

\bibitem{Gomez} X. Gomez,  \textit{ Classification of three-dimensional Lie
bialgebras}, J. Math. Phys. \textbf{ 41} (2000) 4939.

\bibitem{Snobl} L. Hlavaty and L. Snobl, \textit{ Classification of 6-dimensional Manin
triples}, \texttt{[arXiv:math-ph/0202209]}.

\bibitem{RHR} A. Rezaei-Aghdam, M. Hemmati, A. R. Rastkar, \textit{
classification of real three-dimensional Lie bialgebras and their
Poisson-Lie groups}, \textit{ J. Phys. A:Math.Gen.} \textbf{ 38}
(2005) 3981-3994, \texttt{[arXiv:math-ph/0412092]}.

\bibitem{Kosmann} Y. Kosmann-Schwarzbach, \textit { Lie bialgebras,
Poisson-Lie groups and dressing transformations integrability of
nonlinear Systems}: Proc.(Pondicherry) ed Y Kosmann-Schwarzbach, B
Grammaticos and K M Tamizhmani (Berlin: Springer, 1996) pp 104-70.

\bibitem{N.A}  N. Andruskiewitsch, \textit{Lie Superbialgebras and Poisson-Lie supergroups},
 Abh. Math. Sem. Univ. Hamburg {\bf 63} (1993) 147-163.

\bibitem{Bs}   N. Beisert and E. Spill, \textit {The
classical r-matrices of AdS/CFT and its Lie bialgebras structure},
\texttt{[arXiv:0708.1762 [hep-th]]}.

\bibitem {Geer}  N. Geer, \textit{Etingof-Kazdan quantization of Lie
super bialgebras}, \texttt{[arXiv:math.QA/0409563]}.

\bibitem {J.z}  C. Juszczak and J. T. Sobczyk,  \textit {Classification of low dimentional Lie
super-bialgebras, } \textit{J. Math. Phys.} \textbf{39}, (1998)
4982-4992.


\bibitem{J}  C. Juszczak,  \textit {Classification of $osp(2|2)$ Lie super-bialgebras, }
\texttt{[arXiv:math.QA/9906101]}.

\bibitem{KARAALI}  G. Karaali, \textit{ A New Lie Bialgebra Structure on
sl(2,1)}, \textit{ Contemp. Math.} \textbf{413}  (2006),
pp.101-122.

\bibitem{B}  N. Backhouse, \textit{ A classification of four-dimensional Lie
superalgebras}, \textit{J. Math. Phys.} \textbf{ 19} (1978)
2400-2402.

\bibitem{D}  B. DeWitt, \textit{ Supermanifolds,} Cambridge University Press
1992.

\bibitem{F1} \textit{ Note that the bracket of one boson with one boson or one
fermion is usual commutator but for one fermion with one fermion
is anticommutator. Furthermore we identify grading of indices by
the same indices in the power of (-1), for example
$grading(i)\equiv i$; this is the notation that DeWitt applied in
his book (Ref. \cite{D}).}

\bibitem{F2} \textit{ For the proof of this proposition it is enough to use (11) and
$\delta^\prime(X_i) = (-1)^{jk}{{\tilde{f}}'^{jk}}_{\; \; \; \:i}
X_j \otimes X_k$ and (22) in (5), then if super Jacobi and mixed
super Jacobi identities [(2) and (12)] for ${\tilde{f}^{jk}}_{\;
\; \; \:i}$ are satisfied then they satisfy for
${{\tilde{f}}'^{jk}}_{\; \; \; \:i}$ as well and vise versa.}

\bibitem{RE} A. Eghbali and A. Rezaei-Aghdam, \textit{ Classification
of four-dimensional Lie super-bialgebras of the type $(2 , 2)$},
work in progress.

\bibitem{F3} \textit{ Note that the effect of $A$ is linear and even i.e.
$$
A(a_i^{\;\;j}X_j)=a_i^{\;\;j} A(X_j).
$$}
\bibitem{F4} \textit{ Note that the superalgebra $A$ is one dimensional
Abelian Lie superalgebras with one fermionic generator where Lie
superalgebra $A_{1,1}$ is its bosonization.}

\bibitem{F5} \textit{  Note that in this classification one can omit the $i=\sqrt{-1}$
from commutation relations and to obtain Lie super-bialgebras in
the nonstandard basis.}

\bibitem{K.S1} C. Klim\v {c}ik and P. \v {S}evera, \textit{ Dual non-Abelian duality and
the Drinfeld double}, \textit{Phys. Lett.} \textbf{ B351} (1995)
445-462, \texttt{[arXiv:hep-th/9502122]}.\\
C. Klim\v {c}ik, \textit{ Poisson-Lie $T$-duality}, \textit{Nucl.
Phys. Proc. Suppl.} \textbf{ 46} ( 1996) {116-121},
\texttt{[arXiv:hep-th/9509095]}.

\bibitem {ER}  A. Eghbali, A. Rezaei-Aghdam, \textit
{Poisson-Lie T-dual sigma models on supermanifolds}, JHEP 2009,
{\bf 094}  (2009), \texttt{[arXiv:0901.1592v3 [hep-th]]}.

\bibitem {ER1}  A. Eghbali and A. Rezaei-Aghdam, \textit {Classical $r$-matrices of
two and three dimensional Lie super-bialgebras and their
Poisson-Lie supergroups}, \texttt{[ arXiv:0908.2182 [math-ph]]}.

\bibitem{Von} R. Von Unge, \textit { Poisson-Lie T-plurality},  JHEP, 2002,  {\bf 014} (2002) (preprint
\texttt{[arXiv:hep-th/0205245]}).

\end{thebibliography}
\end{document}